\documentstyle[12pt]{article}
\def\k{{\rm {\bf k}}}
\def\p{{\rm {\bf p}}}
\def\q{{\rm {\bf q}}}

\def\Tr{{\rm Tr}}
\def\lg{{\rm log}}
\def\exp{{\rm exp}}
\def\a{{\rm {\hat a}}}

\begin{document}

\hfill BI-TP 95/13

\hfill March 1995

\vspace{1.5cm}

\begin{center}
{\bf SELF-ENERGY PECULIARITIES OF THE HOT GAUGE THEORY
AFTER SYMMETRY BREAKING }
 \footnote{Research is partially supported by "Volkswagen-Stiftung"}

\vspace{1.5cm}
{\bf O.K.Kalashnikov}
\footnote{Permanent address: Department of
Theoretical Physics, P.N.Lebedev Physical Institute, Russian
Academy of Sciences, 117924 Moscow, Russia. E-mail address:
kalash@td.lpi.ac.ru}

Fakult\"at f\"ur Physik

Universit\"at Bielefeld

D-33501 Bielefeld, Germany

\vspace{2.5cm}
{\bf Abstract}
\end{center}
A tensor representation of the gluon propagator is found within covariant
gauges for a non-Abelian theory after symmetry breaking due to $<A_0>\ne 0$
and the exact equations which determine the dispersion laws of plasma
excitations are explicitly obtained. In the high temperature region and
fixing the Feynman gauge we solved these equations and found the damping
of the plasma oscillations and the shifting of their frequency. The phase
transition of a gauge symmetry restoration is estimated to be $\alpha_c(T)
\approx{4/3}$.

\newpage

\section{Introduction}

Today there is a hope that a non-Abelian gauge theory which acquires a
new vacuum after the global gauge symmetry spontaneously breaking will be
free of infrared divergencies [1,2] and able to give the reliable predictions
for physics. A new vacuum is formed by a nonzero $A_{0}$-condensate
which results from the infrared instability of a non-Abelian gauge theory and
its appearence minimizes the thermodynamical potential beyond the trivial
vacuum (the old scenario proposed many years ago [3]). This scenario was
studied in connection with the deconfinement problem [4,5] where it was shown
that the vacuum with $<A_{0}>\ne0$ breaks a $Z(N)$-symmetry and gauge theory
loses their confining property in the high temperature region. Resently the
scenario with $<A_{0}>\ne0$ was again studied in many papers (on the two-loop
level [6-10] and with the higher order corrections taken into account
[11,12], see also review [13] for details), but unfortunately up to now the
possibility of $A_0$-condensation is not rigorous proved. The problem arises
when the multi-loop corrections are taken into account and also this scenario
is not confirmed by using the partition function found for the eigenvalues
on the Wilson line [14]. Today this problem remains however it is not a
reason to reject this scenario which opens the new possibilities for physics
(the same opinion see in [13] and Ref.[15] is interesting as well). Of course,
the question of gauge dependence of this phenomenon is also not completely
clear. The $A_{0}$-quantity being a minimum of an effective action is
gauge-dependent at the beginning but unfortunately the perturbative
thermodynamical potential (namely its $g^4$-term and the higher terms as well)
are also gauge-dependent [1] even for the case when gauge-covariance of
perturbative calculations are checked by using the Nielsen identity [9].

The goal of this paper is to investigate the self-energy peculiarities of
a non-Abelian gauge theory after symmetry breaking and to verify a
selfconsistency of the scenario with $<A_0>\ne 0$. This theory being
more complicated requires a new formalism for clarifying its sence since
at the beginning the self-energy tensor has the new structures which are
found by using the gauge group indices (in accordance with a classical
field vector). Our formalism is built in a covariant background gauge with
an arbitrary $\alpha$-paramiter and the exact dispersion equations are
explicitly obtained. It is shown that there are the two
singled out gauges (the Landau and Feynman ones) where these equations
have a more simple form and we solved them in the high temperature region
by fixing the Feynman gauge. It is established that the plasma oscillations
(both the transverse and longitudinal ones) have a finite damping everywhere
even in the high temperature limit and the well-known plasmon frequency is
shifted but differently for each vector field. Extending this scenario to
the low temperature region we find the point $T_c$ below which the
$A_0$-condensate should be disappear to keep the squared plasma frequency
positive. The estimation gives that the phase transition is localized
in $\alpha_c(T)\approx 4/3$ and for the lower temperatures the gauge
symmetry is restored.

\section{Tensor structure of the gluon propagator}

Our formalism is built by using the standard Green function technique at
$T\ne 0$ in the background gauge with an arbitrary parameter $\alpha$. To
simplify our calculation we consider only the SU(2)-gauge group and the
external classical field is chosen to be
\setcounter{equation}{0}
\begin{eqnarray}
{\bar A}_{\mu}^a=\delta_{\mu 4}\delta^{a3}\frac{\pi T}{g}x
\end{eqnarray}
Due to this choice there is a new vector in the theory and the self-energy
tensor ( which determines the standard gluon propagator) has the form
\begin{eqnarray}
\Pi_{\mu\nu}^{ab}=\delta^{ab}\Pi_{\mu\nu} + K {\bar A}_{\mu}^{a}
{\bar A}_{\nu}^{b} + R\;(\varepsilon^{afc}{\bar A}_{\mu}^{f})(\varepsilon^{ctb}
{\bar A}_{\nu}^{t})
\end{eqnarray}
So the initial SU(2)- group is broken and the two tensors
$\Pi_{\mu\nu}^{||}(\p_4,\p)$ and $\Pi_{\mu\nu}^{\perp}(p_4,\p)$ should be
considered independly. Both these tensors are not transverse for any covariant
gauges and, in a more general case, there are four scalar functions which
determine them as follows
\begin{eqnarray}
\Pi=\Pi_{t}\cdot A+\Pi_{l}\cdot B+\Pi_{c}\cdot C+\Pi_{d}\cdot D
\end{eqnarray}
where the four tensors are chosen to be [16]
\begin{eqnarray}
A=I-B-D\;\; , \;\; B=\frac{V\circ V}{V^2}\;\; , \;\; C=\frac{Q\circ V+V\circ Q}
{\sqrt{2}Q^2|\q|}\;\; , \;\; D=\frac{Q\circ Q}{Q^2}
\end{eqnarray}
which being in a covariant form are more convenient although
they are only slightly different from the tensors used earlier in [17].
Here $V=Q^2U-(U\cdot Q)Q$, and $U=(1,{\bf 0})$ is the four-velocity of the
thermal bath at rest. As usual $U^2=1$ and $V^2=Q^2|\q|^2$. The algebra of
these
tensors is found as follows
\begin{eqnarray}
&&A^2=A\;\ , \;\;B^2=B\;\; , \;\;D^2=D\;\; , \;\;C^2=\frac{1}{2}(B+D)
\;\; , \;\;\nonumber\\&& AB=AC=AD=BD=0\;\; , \;\;BC=CD=\frac{V\circ Q}
{\sqrt{2}Q^2|\q|}\;\; , \;\;\nonumber\\&& \frac{1}{2}\Tr A^2=\Tr B^2
=\Tr C^2=\Tr D^2=1
\end{eqnarray}
where I is the unit tensor and we use the Euclidean metric $(Q^2=q_4^2
+|\q|^2)$.

For the Feynman gauge (where $\alpha=1$) there is a simplification and
the polarization tensors for both sectors of the broken SU(2)-theory (instead
of (3)) have a more simple structure [1]
\begin{eqnarray}
\Pi = a\cdot (I-D)+\frac{b|\q|^2}{(U\cdot Q)^2}\cdot B+Z\cdot \frac{U\circ U}
{(U\cdot Q)}
\end{eqnarray}
where one can finds the following relations
\begin{eqnarray}
\Pi_{t}&=&a,\\
\Pi_{l}&=&a+\frac{|\q|^2}{(U\cdot Q)^2}(b+Z\frac{(U\cdot Q)}{Q^2});\; , \;\;
\frac{Z\sqrt{2}|\q|}{Q^2} = \Pi_{c}\;\; , \;\;
\frac{Z(U\cdot Q)}{Q^2} = \Pi_{d}\nonumber
\end{eqnarray}
The representation (6) makes the Feynman gauge very convenient for the
practical calculations and below namely this gauge will be used. But here
within the exact algebra we continue to work with Eq.(3) and bearing in mind
the standard definition of the inverse propagator
\begin{eqnarray}
{\cal D}^{-1} = (Q^2+\Pi_{t})\cdot A + (Q^2+\Pi_{l})\cdot B + \Pi_{c}\cdot C +
( \frac{Q^2}{\alpha}+\Pi_{d})\cdot D
\end{eqnarray}
the explicit form of the gluon propagator is found to be
\begin{eqnarray}
{\cal D}&=&\frac{ A}{Q^2+\Pi_{t}}+\frac{1}{(Q^2+\alpha\Pi_{d})
(Q^2+\Pi_{l})-\frac{\alpha}{2}\Pi_{c}^2}\left\{(Q^2+\alpha\Pi_{d})
\cdot B\right.\nonumber\\
&-&\left.\alpha\Pi_{c}\cdot C+\alpha(Q^2+\Pi_{l})\cdot D\right\}
\end{eqnarray}
that is valid for any $\alpha$. In accordance with Eq.(9) there are two poles
\begin{eqnarray}
Q^2+\Pi_{t}=0\,,\qquad(Q^2+\alpha\Pi_{d})(Q^2+\Pi_{l})-\frac{\alpha}{2}
\Pi_{c}^2=0
\end{eqnarray}
which determine the modified dispersion equations for a gauge theory after
the symmetry is spontaneously broken. For an arbitrary gauge the longitudinal
oscillation spectrum is very complicated but there is one singled gauge
$\alpha=0$ where one finds Eqs.(10) in a rather simple form
\begin{eqnarray}
Q^2+\Pi_{t}=0\,,\qquad Q^2+\Pi_{l}=0
\end{eqnarray}
which is very close to the earlier known one. However, the Feynman gauge is
also convenient since in this case the $\Pi_{l}$-function can be easily found
from the self-energy diagram representation (through the $\Pi_{44}$-function).
This equality is established to be
\begin{eqnarray}
\Pi_{l}=\frac{Q^2}{|\q|^2}(\Pi_{44}-\frac{Z}{(Q\cdot U)})+\frac{Z|\q|^2}
{(Q\cdot U)Q^2}
\end{eqnarray}
and using Eq.(7) one can transform the dispersion equation for the longitudinal
oscillations to the form
\begin{eqnarray}
(Q^2+\frac{Z(Q\cdot U)}{Q^2})\left\{1+\frac{1}{|\q|^2}(\Pi_{44}
-\frac{Z}{(Q\cdot U)})\right\}+\frac{Z|\q|^2}{Q^2(Q\cdot U)}=0
\end{eqnarray}
which for $|\q|^2=0$ reproduces a rather simple equation
\begin{eqnarray}
1+\frac{1}{|\q|^2}(\Pi_{44}-\frac{Z}{(Q\cdot U)})=0
\end{eqnarray}
that allows to find the energy gap in the spectrum of the longitudinal
oscillations with the aid of the simple calculation.

\section{The high temperature spectra of plasma oscillations}

Here we calculate the high temperature spectra of the plasma oscillations
in the Feynman gauge (where $\alpha=1$) and discuss their properties.
Since the broken SU(2)-theory has two different sectors (named as
"longitudinal" and "transverse" in accordance with its gauge group indices)
there are two different expressions for each scalar function in Eq.(6).

\subsection{Transverse oscillations in the SU(2)-transverse \protect\\ sector}

The SU(2)-transverse sector contains two gauge fields $V_{\mu}^{\pm}=(V_{\mu}^1
\mp iV_{\mu}^2)/\sqrt{2}$ (with their ghost ones) and after the gauge symmetry
breaking a new vacuum redefines the $p_4$-dependence for all functions within
this sector. Now all momenta are shifted ( ${\hat p}_4=p_4+\mu$ where
$\mu=\pi Tx$) and the polarization tensor is not transverse and has a structure
like (6).

In the one-loop approximation this tensor is found to be [1]
\begin{eqnarray}
-\Pi_{\mu\nu}^{\perp}({\hat p}_4,\p)&=&\frac{1}{\beta}\sum_{k_4}
\int\frac{d^3\k}{(2\pi)^3}
\left\{\frac{g^2}{k^2(k+{\hat p})^2}[\delta_{\mu\nu}(2k^2+5{\hat p}^2
+2k{\hat p})\right.\\
&+&\left.8k_{\mu}k_{\nu}-2{\hat p}_{\mu}{\hat p}_{\nu}+4(k_{\mu}{\hat p}_{\nu}
+k_{\nu}{\hat p}_{\mu})]-3\delta_{\mu\nu}g^2(\frac{1}{{\hat k}^2}
+\frac{1}{k^2})\right\}\nonumber
\end{eqnarray}
where ${\hat p}=({\hat p}_4,\p)$ and ${\hat p}_4=(p_4+\mu)$ as it is mentioned
above.

At first we calculate the $\Pi_{44}^{\perp}({\hat p}_4,\p)$-function
within Eq.(15). The standard integrals are used and after all algebra being
performed one finds that this function has the form
\begin {eqnarray}
&&\Pi_{44}^{\perp}({\hat p}_4,\p)=\frac{g^2N}{2\pi^2}\int\limits_{0}^{\infty}
d|\k|\left\{|\k|(\frac{n^++n^-}{2}+n)\right.\nonumber\\
&&\hspace{0cm}\left.+\frac{1}{8|\p|}\Biggr[({\hat p}_4^2+2\p^2-4\k^2)
\Bigr[(\frac{n^++n^-}{2}+n)\lg(\a^+\a^-)-\frac{n^+-n^-}{2}
\lg(\frac{\a^+}{\a^-})\Bigr]\right.\nonumber\\
&&\hspace{0cm}\left.+4|\k|(i{\hat p}_4)
\Bigr[(\frac{n^++n^-}{2}+n)\lg(\frac{\a^+}{\a^-})-\frac{n^+-n^-}{2}
\lg(\a^+\a^-)\Bigr]\Biggr]\right\}
\end{eqnarray}
where $n=\left\{\exp[\beta|\k|]-1\right\}^{-1}$ and all other abreviations are
introduced to be
\begin{eqnarray}
n^{\pm}&=&\left\{\exp[\beta(|\k|\pm i\mu)]-1\right\}^{-1}\nonumber\\
\a^{\pm}&=&\frac{(\p^2+{\hat p}_4^2-2|\k|\|\p|)\pm 2i|\k|{\hat p}_4}
{(\p^2+{\hat p}_4^2+2|\k||\p|)\pm 2i|\k|{\hat p}_4}
\end{eqnarray}
In the leading order of T (for the high temperature region) the integral in
Eq.(16) is calculated completely and the simple expression arises
\begin{eqnarray}
\Pi_{44}^{\perp}({\hat p}_4,|\p|)&=&\frac{g^2T^2N}{\pi^2}
[I_1(\frac{x}{2})+I_1(0)]\left\{1
-\frac{\xi}{2}\lg\frac{\xi+1}{\xi-1}\right\}\nonumber\\
&+&i\frac{g^2T^3N}{4\pi^2|\p|}I_2(\frac{x}{2})\lg\frac{\xi+1}{\xi-1}
\end{eqnarray}
where $\xi={i{\hat p}_4}/|\p|$ and the integrals $I_i$ being treated in the
dimensionless variable $|\k|/T$ are found to be
\begin{eqnarray}
I_1(\frac{x}{2})&=&\int\limits_{0}^{\infty}zdz\frac{n^++n^-}{2}
=\pi^2B_2(\frac{x}{2})\nonumber\\
I_2(\frac{x}{2})&=&i\int\limits_{0}^{\infty}z^2dz(n^+-n^-)=\frac{(2\pi)^3}{3}
B_3(\frac{x}{2})
\end{eqnarray}
Here $x=\mu/\pi T$ and $B_i(z)$ are the standard Bernoulli polynomials
\begin{eqnarray}
B_2(z)=z^2-|z|+1/6\;\; , \;\;B_3(z)=z^3-3\varepsilon(z)z^2/2+z/2
\end{eqnarray}
with $\varepsilon(z)=z/|z|$ and $\varepsilon(0)=0$.

To calculate the $\Pi_{t}^{\perp}({\hat p}_4,\p)$-function we use the
convenient
formula
\begin{eqnarray}
\Pi_{t}=\frac{1}{2}\left\{\sum_{i}\Pi_{ii}-\frac{p_4^2}{\p^2}(\Pi_{44}
-\frac{Z}{p_4})\right\}
\end{eqnarray}
which holds in the Feynman gauge in accordance with Eq.(6).
We again return to Eq.(15) and use the formulas obtained above to perform the
necessary algebra. The result has the form
\begin{eqnarray}
&\!\!\!\!\!\!\!\!\!\!\!&\Pi_{t}^{\perp}({\hat p}_4,|\p|)=\frac{g^2N}{4\pi^2}
\int\limits_{0}^{\infty}
d|\k|\left\{|\k|\Bigr[(1-\frac{{\hat p}_4^2}{\p^2})(\frac{n^++n^-}{2}+n)
+\frac{i|\k|{\hat p}_4}{\p^2}
(n^+-n^-)\Bigr]\right.\nonumber \\
&\!\!\!\!\!\!\!\!\!\!\!&+\left.\frac{p^2}{8|\p|^3}\Biggr[(3\p^2
-{\hat p}_4^2+4\k^2)\Bigr[(\frac{n^++n^-}{2}+n)\lg(\a^+\a^-)
-\frac{n^+-n^-}{2}\lg(\frac{\a^+}{\a^-})\Bigr]\right.\nonumber \\
&\!\!\!\!\!\!\!\!\!\!\!&-\left.4|\k|(i{\hat p}_4)
\Bigr[(\frac{n^++n^-}{2}+n)\lg(\frac{\a^+}{\a^-})-\frac{n^+-n^-}{2}\lg(\a^+\a^-)
\Bigr]\Biggr]\right\}
\end{eqnarray}
where $p^2=\p^2+{\hat p}_4^2$ and we take into account that
\begin{eqnarray}
\frac{Z^{\perp}}{{\hat p}_4}=i\frac{g^2T^3N}{2\pi^2(i{\hat p}_4)}
I_2(\frac{x}{2})
\end{eqnarray}
In the high temperature region Eq.(22) is simplified to be
\begin{eqnarray}
\Pi_{t}^{\perp}({\hat p}_4,|\p|)&=&\frac{g^2T^2N}{2\pi^2}(\xi^2-1)
[I_1(\frac{x}{2})+I_1(0)]\left\{\frac{\xi^2}
{\xi^2-1}-\frac{\xi}{2}\lg\frac{\xi+1}{\xi-1}\right\}\nonumber\\
&+&i\frac{g^2T^3N}
{8\pi^2|\p|}(\xi^2-1)I_2(\frac{x}{2})\left\{\lg\frac{\xi+1}{\xi-1}
-\frac{2\xi}{\xi^2-1}
\right\}
\end{eqnarray}
where $\xi={i{\hat p}_4}/|\p|$. Now the dispersion equation
(${\hat p}_4^2+\p^2+\Pi_{t}^{\perp}=0$ ) for transverse plasma oscillations
in the high temperature region has the form
\begin{eqnarray}
(i{\hat p}_4)^2&=&\frac{g^2T^2N}{2\pi^2}[I_1(\frac{x}{2})+I_1(0)]\;\xi^2
\left\{\frac{\xi^2}{\xi^2-1}
-\frac{\xi}{2}\lg\frac{\xi+1}{\xi-1}\right\}\nonumber\\
&+&i\frac{g^2T^3N}
{8\pi^2|\p|}I_2(\frac{x}{2})\;\xi^2\left\{\lg\frac{\xi+1}{\xi-1}
-\frac{2\xi}{\xi^2-1}\right\}
\end{eqnarray}
Eq.(25) has real and imaginary parts and determines both the spectrum
of plasma oscillations and their damping. It is a rather complicated equation
and we solve it only in the long-wave length limit for $\xi\rightarrow \infty$
where Eq.(25) is reduced to be
\begin{eqnarray}
(i{\hat p}_4)^2=\frac{g^2T^2N}{3\pi^2}[I_1(\frac{x}{2})+I_1(0)]
-i\frac{g^2T^3N}{6\pi^2(i{\hat p}_4)}I_2(\frac{x}{2})
\end{eqnarray}
Solving Eq.(26) we consider that $ip_4=\omega=\Delta+i\Gamma$ and find
two simple equations
\begin{eqnarray}
\Delta^2-(\Gamma+\mu)^2=\frac{K(\Gamma+\mu)}{\Delta^2+(\Gamma+\mu)^2}+\Lambda^2
\;\; , \;\;2(\Gamma+\mu)=\frac{K}{\Delta^2+(\Gamma+\mu)^2}
\end{eqnarray}
where the new abreviations are:
\begin{eqnarray}
\Lambda^2=\frac{g^2T^2N}{3\pi^2}[I_1(\frac{x}{2})+I_1(0)]\;\; , \;\;K=
-\frac{g^2T^3N}{6\pi^2}I_2(\frac{x}{2})
\end{eqnarray}
For the trivial vacuum (where $x=0$) $\Lambda^2=\omega_{pl}^2=g^2T^2N/9$
and  $K=0$. Eq.(27) can be simplified and solved approximately for small
$g^2$ where $4(\Gamma^2+\mu)<<\Lambda^2$. This solution has the form
\begin{eqnarray}
\Delta^2=3(\Gamma+\mu)^2+\Lambda^2\;\; , \;\;2(\Gamma+\mu)=K/\Lambda^2
\end{eqnarray}
and for the physical vacuum where $\mu=\pi Tx$ and $x=g^2/4\pi^2$ [1]
one finds (if $g^2<<1$) that
\begin{eqnarray}
(\Gamma+\mu)\approx-\frac{g^2T}{8\pi}\;\; , \;\;\omega^2\approx
3(\frac{g^2T}{8\pi})^2+\frac{g^2T^2N}{9}(1-\frac{3g^2}{8\pi^2})
\end{eqnarray}
Studying Eq.(30) one can see that there is a point $T_c$ where $\omega^2=0$.
At this point the phase transition occurs and for all $T\le T_c$ the gauge
symmetry is restored. To etimate $T_c$ we solve the equation $\omega^2=0$
and find that $\alpha_c(T)=g_c^2(T)/4\pi^2\approx 4/3$ for the SU(2)-group.
This is our main result for this setion but we would also like to stress that
the same spectrum  as (30) arises from Eq.(14) which determines the long-wave
length limit for the longitudinal plasma oscillations. Indeed, using Eqs.
(18) and (23) one finds that
\begin{eqnarray}
(i{\hat p}_4)^2&=&\;\xi^2\;\left\{i\frac{g^2T^3N}{2\pi^2}I_2(\frac{x}{2})
\;[\frac{1}{(i{\hat p}_4)}-
\frac{1}{2|\p|}\lg\frac{\xi+1}{\xi-1}]\right.\nonumber\\
&-&\left.\frac{g^2T^2N}{\pi^2}[I_1(\frac{x}{2})+I_1(0)]\;[1-\frac{\xi}{2}
\lg\frac{\xi+1}{\xi-1}]\right\}
\end{eqnarray}
and then taking the limit $\xi\rightarrow\infty$ within Eq.(31) we easily
establish that Eq.(25) is exactly reproduced.
This means that at the point $|\p|=0$ the spectra of transverse and
longitudinal oscillations coincide and determine the spectrum of the unique
excitation with the three degrees of freedom.

\subsection{Transverse oscillations in the SU(2)-longitudinal sector}

The londgitudinal sector of the SU(2)-theory  contains one real gauge field
$V_\mu^3$ (with its ghost one) and a new vacuum keeps the usual
$p_4$-dependence (the own momenta are not shifted). Nevertheless the
$\Pi_{\mu\nu}^{||}(p_4.\p)$-tensor is not transverse the same as in the
previous
case and has a structure like (6).

In the one-loop approximation this tensor is found to be [1]
\begin{eqnarray}
-\Pi_{\mu\nu}^{||}(p_4,\p)&=&\frac{1}{\beta}\sum_{k_4}
\int\frac{d^3\k}{(2\pi)^3}
\left\{\frac{g^2}{{\hat k}^2({\hat k}+p)^2}[\delta_{\mu\nu}(2{\hat k}^2+5p^2
+2{\hat k}p)\right.\nonumber\\
&+&\left.8{\hat k}_{\mu}{\hat k}_{\nu}-2p_{\mu}p_{\nu}+4({\hat k}_{\mu}p_{\nu}
+{\hat k}_{\nu}p_{\mu})]-\frac{6\delta_{\mu\nu}g^2}{{\hat k}^2}\right\}
\end{eqnarray}
where ${\hat k}=({\hat k}_4,\k)$ and ${\hat k}_4=(k_4+\mu)$. Here we calculate
its scalar functions in accordance with Eq.(6) but below many
details of these calculations will be omitted since they can be found in the
previous section.

The result of this calculations is:
\begin{eqnarray}
&&\Pi_{t}^{||}(p_4,|\p|)=\frac{g^2N}{2\pi^2}\int\limits_{0}^{\infty}d|\k|
\left\{|\k|\Bigr[(1-\frac{p_4^2}{\p^2})\frac{n^++n^-}{2}+\frac{i|\k|p_4}{\p^2}
(n^+-n^-)\Bigr]\right.\nonumber\\
&&\hspace{0cm}\left.+\frac{p^2}{8|\p|^3}\Biggr[(3\p^2-p_4^2+4\k^2)
\Bigr[\frac{n^++n^-}{2}\lg(a^+a^-)-\frac{n^+-n^-}{2}\lg(\frac{a^+}{a^-})
\Bigr]\right.\nonumber\\
&&\hspace{0cm}\left.-4|\k|(ip_4)
\Bigr[\frac{n^++n^-}{2}\lg(\frac{a^+}{a^-})-\frac{n^+-n^-}{2}\lg(a^+a^-)
\Bigr]\Biggr]\right\}
\end{eqnarray}
where all abreviations are the same as previously. Here $a^{\pm}$ repeats
Eq.(17) where ${\hat p}_4$ is replaced by $p_4$.
In the high temperature region (if $p_4\ne 0$ and $\xi={ip_4}/|\p|$) one
finds that Eq.(33) is simplified to be
\begin{eqnarray}
\Pi_{t}^{||}(p_4,|\p|)&=&\frac{g^2T^2N}{\pi^2}(\xi^2-1)
I_1(\frac{x}{2})\left\{\frac{\xi^2}
{\xi^2-1}-\frac{\xi}{2}\lg\frac{\xi+1}{\xi-1}\right\}\nonumber\\
&+&i\frac{g^2T^3N}
{4\pi^2|\p|}(\xi^2-1)I_2(\frac{x}{2})\left\{\lg\frac{\xi+1}{\xi-1}
-\frac{2\xi}{\xi^2-1}
\right\}
\end{eqnarray}
and the high temperature dispersion equation for the transverse plasma
oscillations has the form
\begin{eqnarray}
(ip_4)^2&=&\frac{g^2T^2N}{\pi^2}\;I_1(\frac{x}{2})\;\xi^2\left\{\frac{\xi^2}
{\xi^2-1}-\frac{\xi}{2}\lg\frac{\xi+1}{\xi-1}\right\}\nonumber\\
&+&i\frac{g^2T^3N}
{4\pi^2|\p|}I_2(\frac{x}{2})\;\xi^2\left\{\lg\frac{\xi+1}{\xi-1}
-\frac{2\xi}{\xi^2-1}\right\}
\end{eqnarray}
The same as previously Eq.(35) has real and imaginary parts and determines
both the spectrum of plasma oscillations and their damping. It is a
complicated equation as well and we solve it only in the long-wave length
limit for $\xi\rightarrow \infty$ where Eq.(35) is reduced to be
\begin{eqnarray}
(ip_4)^2=\frac{2g^2T^2N}{3\pi^2}I_1(\frac{x}{2})
-i\frac{g^2T^3N}{3\pi^2(ip_4)}I_2(\frac{x}{2})
\end{eqnarray}
Here we consider that $ip_4=\omega=\Delta+i\Gamma$ and find two simple
equations
\begin{eqnarray}
\Delta^2-\Gamma^2=\frac{K\Gamma}{\Delta^2+\Gamma^2}+\Lambda^2
\;\; , \;\;2\Gamma=\frac{K}{\Delta^2+\Gamma^2}
\end{eqnarray}
where the new abreviations are:
\begin{eqnarray}
\Lambda^2=\frac{2g^2T^2N}{3\pi^2}I_1(\frac{x}{2})\;\; , \;\;K=
-\frac{g^2T^3N}{3\pi^2}I_2(\frac{x}{2})
\end{eqnarray}
For the trivial vacuum (where $x=0$) $\Lambda^2=\omega_{pl}^2=g^2T^2N/9$
and  $K=0$ the same as previously.
Eq.(37) can be simplified and solved approximately for small $g^2$ where
$4\Gamma^2<<\Lambda^2$. This solution has the form
\begin{eqnarray}
\Delta^2=3\Gamma^2+\Lambda^2\;\; , \;\;2\Gamma=K/\Lambda^2
\end{eqnarray}
and for the physical vacuum where $x=g^2/4\pi^2$ [1] one finds (if $g^2<<1$)
that
\begin{eqnarray}
\Gamma\approx-\frac{g^2T}{4\pi}\;\; , \;\;\omega^2\approx
3(\frac{g^2T}{4\pi})^2+\frac{g^2T^2N}{9}(1-\frac{3g^2}{8\pi^2})
\end{eqnarray}
In the end of this setion we again stress that the same spectrum as (40)
arises from Eq.(14) which determines the long-wave length limit for the
longitudinal plasma oscillations.

\section{Conclusion}

To summarize we have established a finite damping of plasma oscillations and
the shifting of their frequency for the broken SU(2)-gauge theory when
$A_0$-condensation takes place.  This condensate occurs at the enough high
temperature and disappears when the temperature falls below $T_c$: a
temperature for which the plasma frequency of transverse oscillations in the
SU(2)-transverse sector (see Eq.(30)) becomes equal to zero (here the running
constant $\alpha_c(T)=g_c^2(T)/{4\pi^2}\approx4/3$). Below $T_c$ the gauge
symmetry restores and the theory again
acquires the confining property. Our calculations are performed in the high
temperature limit however we force to fix the Feynman gauge at the beginning
to simplify many expressions and the algebraic transformations. Since we know
[1] that in the used approximation  the thermodynamical potential is
gauge-dependent we do not exclude that the one-loop results obtained for a
self-energy are gauge-dependent as well although this dependence is not
found explicitly. However we establish that there are two gauges (the Landau
and Feynman ones) which are singled out for this task and the longitudinal
spectrum of plasma oscillations essentially depends on this choice. Since
this dependence arises on the agebraic level for the exact tensors there are
no reasons to wait that the situation changes when the multi-loop corrections
will be taken into account. Today the problem is to understand  whether
this dependence has a qualitative character or it leads only to a quantitative
changes the physical results found within this theory. Of course, it is
desirable to find a nonperturbative approach which allows to build the
gauge-independent (or practically gauge-independent) thermodynamical potential.
Within any perturbative calculations there is no chance to make the broken
theory with $<A_0>\ne 0$ to be gauge-independent although we can demonstrate
that any of these approximations are gauge-covariant and self-consistent on the
level of general identities (see [9] for the details).

\begin{center}
{\bf Acknowledgements}
\end {center}

I would like to thank Rudolf Baier for useful discussions as well as
all the colleagues from the Department of Theoretical Physics of the
Bielefeld University for the kind hospitality.

\newpage

\begin{center}
{\bf References}
\end{center}

\renewcommand{\labelenumi}{\arabic{enumi}.)}
\begin{enumerate}

\item{  R. Anishetty, J. Phys.{\bf G10} (1984) 423.}

\item{ O.K.Kalashnikov, Prog.Theor.Phys.{\bf 92} (1994) 1207.}

\item{ O.K.Kalashnikov, V.V.Klimov, E.Casado, Phys. Lett.{\bf B114}
(1982) 49.}

\item{ N.Wess, Phys. Rev. {\bf D24} (1981) 475; {\bf D25} (1982) 2667.}

\item{ R. Anishetty, J. Phys.{\bf G10} (1984) 439.}

\item{ V.M.Belyaev, V.L.Eletsky, Z.Phys. {\bf C45} (1990) 355.}

\item{ K.Enqvist, K.Kajantie, Z.Phys. {\bf C47} (1990) 291.}

\item{ V.M.Belyaev, Phys. Lett. {\bf B254} (1991) 153.}

\item{ V.V.Skalozub, Mod. Phys. Lett. {\bf A7} (1992) 2895.}

\item{ O.K.Kalashnikov, Phys. Lett. {\bf B302} (1993) 453.}

\item{ V.M.Belyaev, Phys. Lett. {\bf B241} (1990) 91.}

\item{ O.K.Kalashnikov, JETP Lett. {\bf 57} (1993) 773.}

\item{ O.A.Borisenko, J.Boh\'a\v cik, V.V.Skolozub, hep-ph/9405208.}

\item{ A.Gocksch, R.D.Pisarski, Nucl. Phys. {\bf B402} (1993) 657.}

\item{ K.Sailer, A.Sch\"afer, W.Greiner, Preprint KLTE - DTP /1995/1,
 hep-ph/9502234}

\item{ F.Flechsig and H.Schulz, DESY preprint 94-240 (1994),
ITP-UH 20/94, hep-ph/9501229}

\item{ K.Kajantie and J.Kapusta, Ann. Phys. {\bf 160} (1985) 477.}

\end{enumerate}
\end{document}